\documentclass[aps,prl,twocolumn,superscriptaddress,raggedbottom,showpacs,floatfix]{revtex4}

\usepackage{amsmath,amsfonts,amssymb,amsthm,graphics}
\usepackage[next]{inputenc}
\usepackage[dvips]{epsfig}
\usepackage[colorlinks=true,citecolor=blue,linkcolor=blue]{hyperref}
\usepackage{bbm}
\usepackage{bbold}
\usepackage{booktabs}
\usepackage{multirow}
\usepackage{hhline}
\usepackage{amssymb}
\usepackage{comment}

\usepackage{bm}
\usepackage{color}
\usepackage{graphicx,latexsym}
\usepackage{epstopdf}

\begin{document} 
\renewcommand{\vec}{\mathbf}
\renewcommand{\Re}{\mathop{\mathrm{Re}}\nolimits}
\renewcommand{\Im}{\mathop{\mathrm{Im}}\nolimits}

\title{Non-Markovian quantum friction of bright solitons in superfluids}
\author{Dmitry K. Efimkin}
\affiliation{Joint Quantum Institute and Condensed Matter Theory Center, Department of Physics, University of Maryland, College Park, Maryland 20742-4111, USA}

\author{Johannes Hofmann}
\affiliation{Joint Quantum Institute and Condensed Matter Theory Center, Department of Physics, University of Maryland, College Park, Maryland 20742-4111, USA}

\affiliation{T.C.M. Group, Cavendish Laboratory, University of Cambridge, Cambridge CB3 0HE, United Kingdom}

\author{Victor Galitski}
\affiliation{Joint Quantum Institute and Condensed Matter Theory Center, Department of Physics, University of Maryland, College Park, Maryland 20742-4111, USA}

\begin{abstract}
We explore the quantum dynamics of a bright matter-wave soliton  in a quasi-one-dimensional bosonic superfluid with attractive interactions. Specifically, we focus on the dissipative forces experienced by the soliton due to its interaction with Bogoliubov excitations.  Using the collective coordinate approach and the Keldysh formalism, a Langevin equation of motion for the soliton is derived from first principles.  The equation contains a stochastic Langevin force (associated with quantum noise) and  a non-local in time dissipative force, which appears due to inelastic scattering of Bogoliubov quasiparticles off of the moving soliton. It is shown that Ohmic friction (i.e., a term proportional to the soliton's velocity) is absent in the integrable setup. However, the Markovian approximation gives rise to the Abraham-Lorentz force (i.e., a term proportional to the derivative of the soliton's acceleration), which is known from classical electrodynamics of a charged particle interacting with its own radiation. These Abraham-Lorentz equations famously contain a fundamental causality paradox, where the soliton/particle interacts with excitations/radiation originating from future events. We show, however, that the causality paradox is an artifact of the Markovian approximation, and our exact non-Markovian dissipative equations give rise to physical trajectories. We argue that the quantum friction discussed here should be observable in current quantum gas experiments.
 \end{abstract}
\pacs{67.85.De, 67.85.Lm, 03.75.Lm}
\maketitle

In recent years, solitons and soliton-like textures have been subject of much interest and research in quantum superfluids~\cite{Abdullaev08} and optical fibers~\cite{Kivshar1998, Drummond1993}. The hallmark features of solitons are their remarkable robustness and stability, which stem from the integrability of the underlying non-linear model. However, solitons in realistic environments experience dissipative forces and eventually decay. Their quasiclassical motion can be well described by Newton's equation~\cite{Konotop2004}
\vspace{-0.05 cm} 
\begin{equation}
M_{\rm eff} \ddot{X} = - \partial_{X} U  + F[X(t)] ,
\end{equation}
\vspace{-0.05 cm} 
where $X$ is the soliton's position, $M_{\rm eff}$ is its effective mass, and the right-hand  side contains external forces due to a confining potential $U$ and the friction force, $F[X(t)] $.   The standard Ohmic friction with $F[{X}] = - \gamma \dot{X}$ has been considered before by many researchers~\cite{Fedichev1999, Muryshev2002, Sinha2006, Gangardt2010, Schecter2012, Wadkin2012, Jackson2007, Cockburn2010,Efimkin2015}. It can be shown that the friction coefficient $\gamma$ is proportional to an integral of the reflection coefficients of the Bogoliubov excitations off the soliton~\cite{Fedichev1999}. However, there is a non-trivial caveat in pristine integrable setups, where the soliton represents a  {\em reflectionless} potential for excitations. Hence, if integrability is preserved, Ohmic friction is strictly absent. This has motivated the authors of Refs.~\cite{Muryshev2002, Sinha2006, Gangardt2010, Schecter2012} to introduce integrability-breaking terms to induce a non-zero friction coefficient $\gamma$. 
 
\begin{figure}[t]
\includegraphics{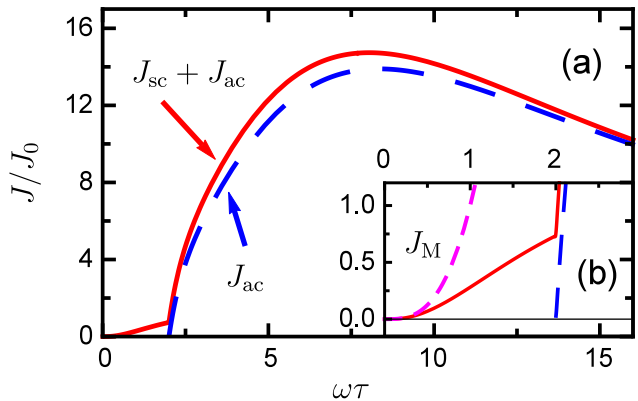}
\caption{(Color online) a) Spectral function of the bath formed by Bogoliubov quasiparticles $J(\omega)=J_\mathrm{sc}(\omega)+J_\mathrm{ac}(\omega)$ (red) and the contribution of creation/annihilation processes $J_\mathrm{ac}(\omega)$ to it (dashed blue). b) The low-frequency part of (a) with the asympotics $J_\mathrm{M}(\omega)$ (dashed purple) given by (\ref{eq:SpectralJMarkovian}). $J_0$ is a constant defined after Eq.~\eqref{eq:SpectralJMarkovian}.} \label{Fig:sf}
\vspace{-0.3 cm}
\end{figure}

Here we revisit this question of soliton dissipation in superfluids, and ask: are there dissipative forces acting on a moving soliton in the perfectly integrable model?  Na{\"\i}vely, by the above argument there should be none because the soliton appears blind to the surrounding cloud of Bogoliubov excitations. We show that this is not the full story and intrinsic, albeit non-Ohmic, friction does exist even if integrability is not broken. It turns out that this problem has a distant cousin in electrodynamics: if a charged particle is moving in an external potential, it is accelerated by the potential and loses energy by emitting electromagnetic radiation. The corresponding classical equation of motion (EOM) contains the Abraham-Lorentz force, ${\bf F}_{\rm AL} \sim \dddot{\bf R}$, giving rise to a famous paradox - the solutions to Abraham-Lorentz equations violate causality (see Refs.~[\onlinecite{Abraham1903, Lorentz1904, LandauLifshitzII,Feynman1945, Gralla2009, Higuchi2004, Aguirregabiria2006, Harvey1991,  Ford1998, Johnson2002}] for a modern point of view and historical perspectives). In this work, we show that the problem of a moving soliton is similar and contains both the Abraham-Lorentz paradox and its resolution by accounting for retardation effects~\cite{Harvey1991, Ford1998, Johnson2002}. There are two key processes, which contribute to intrinsic soliton friction: emission of quasiparticles when accelerating (for dark solitons moving in the presence of gapless phonons) and inelastic scattering of emitted or thermal quasiparticles. In this work, we focus specifically on the simpler case of bright solitons \cite{Khaykovich2002, Strecker2002, Becker2008, Abdullaev08}, for which only the latter mechanism plays out. Our main result - the  soliton's EOM  - is presented below:
\vspace{-0.05 cm} 
\begin{equation}
M \ddot{X}(t) + \int_{0}^t dt' \eta(t-t') \dot{X}(t') = - \partial_{\mathrm{X}} U + f_\mathrm{s}(t), \label{eq:EqOfMotion}
\end{equation}
\vspace{-0.05 cm} 
where  $\eta(t)$ is the dissipation kernel  and $f_\mathrm{s}(t)$ is a stochastic Langevin force.  Its correlation function, $C_\mathrm{s}(t) =\langle f_\mathrm{s}(t) f_\mathrm{s}(0)\rangle$, and $\eta(t)$ are expressed through the same spectral function and fluctuation-dissipation theorem
\vspace{-0.05 cm} 
\begin{align}
\label{eq:FrictionKernel}
\eta(t) &= \frac{2}{\pi}\int_0^\infty d\omega \, \frac{J(\omega)}{\omega} \cos \left(\omega t\right), \\
\label{eq:FrictionKernel2}
C_\mathrm{s}(t) & = \frac{2\hbar }{\pi}\int_0^\infty d\omega \, J(\omega) \coth \left( \frac{\hbar \omega}{2 T} \right) \cos \left(\omega t\right).
\end{align}
\vspace{-0.1 cm} 
An analytic expression for the spectral function of Bogoliubov excitations  is derived below in Eq.~(\ref{eq:SpectralJ}) (see also Fig.~\ref{Fig:sf}).  The Markovian limit (i.e., where $\eta(t-t')$ is approximated by a local-in-time delta-function or its derivatives) of the dissipation force in Eq.~(\ref{eq:EqOfMotion}) contains no Ohmic friction, but gives rise to the Abraham-Lorentz-type force, $F[X] \sim \dddot{X}$ and non-causal soliton trajectories.  However, solutions of Eq.~(\ref{eq:EqOfMotion}) with the full non-Markovian dissipative force contain no causality paradox. 

Our starting point is a $(1+1)$-dimensional field theory, describing a Bose gas with attraction:
\begin{equation}
L = \int dx \left[\phi^* i \hbar \partial_t \phi - \frac{\hbar^2}{2m} |\nabla \phi|^2+\mu |\phi|^2-\frac{g_1}{2} |\phi|^4 \right] , \label{eq:lagrangian}
\end{equation}
where $m$ is the mass of the atoms and $\mu<0$ is the chemical potential. In the context of realistic (quasi)-one-dimensional experiments, the interaction parameter $g_1 = 2 \pi \hbar ^2 a/ m l_{\perp}^2$, where $a$ is the 3D scattering length, $l_{\perp}=\sqrt{\hbar/m\omega_\perp}$, and $\omega_\perp$ is the transverse harmonic confinement frequency~\cite{olshanii98}. 

Importantly, in both 1D and quasi-1D, the interaction part scales as $1/L$ with system size ($L$) and balances the kinetic energy ($\sim 1/L^2$). This implies that  (in contrast to higher dimensions), the attractive Bose gas in one dimension is stable against collapse~\cite{Abdullaev08, Kagan1997}. The attractive nonlinear mean-field interaction energy favors aggregation of particles and counteracts the dispersion of the wavepacket. This leads to the formation of a  {\it bright soliton}, where a Bose-Einstein condensate is localized in a lump of matter with a size set by the coherence length $\xi = \hbar/\sqrt{2m|\mu|}$. The bright soliton solution $\phi_0(x)$ is obtained by minimizing the Lagrangian~\eqref{eq:lagrangian}, i.e., it solves the Gross-Pitaevskii equation. For a soliton with $N$ particles, the wavefunction is given by~\cite{pethick08, PitaevskiiBook}
\begin{align}
\phi_0(x) &= \sqrt{\frac{N}{2 \xi}} \, e^{i\theta} \mathrm{sech}\left(\frac{x-X}{\xi}\right). \label{eq:condensatewf}
\end{align}
Here, $\theta$ and  $X$ are the phase and the coordinate of the  soliton. Note that the energy of a static soliton is independent of $\theta$ and  $X$. 
We consider small-amplitude fluctuations on top of the soliton background, and write the field $\phi(x,t) = \phi_0(x) + \delta \phi(x,t)$, where the soliton wavefunction $\phi_0$ is defined in Eq.~\eqref{eq:condensatewf}. The linear correction vanishes, as $\phi_0$ solves the Gross-Pitaevskii equation. The quadratic correction to the Lagrangian is
$\delta L = 1/2 \int dx \, \Psi^\dagger \left[i \hbar \sigma_3 \partial_t - K_\mathrm{BdG} \right]\Psi $,
where we define $\Psi = (\delta \phi, \delta \phi^*)^T$ and $K_\mathrm{BdG}$ is the positive semidefinite Bogoliubov-de Gennes (BdG) kernel: 
\begin{equation}
K_{\rm BdG} =
\begin{pmatrix}
- \frac{\hbar^2 \nabla^2}{2m} - \mu + 2 g_1 |\phi_0|^2 & g_1 \phi_0^2 \\
g_1 \phi_0^{*2} & - \frac{\hbar^2 \nabla^2}{2m} - \mu + 2 g_1 |\phi_0|^2
\end{pmatrix} .
\end{equation}
The diagonalization proceeds in a similar way as for trapped BECs~\cite{blaizot86,lewenstein96,pethick08}: finite-energy excitations solve the BdG equation $K_{\rm BdG} |k\rangle =\sigma_3 \varepsilon_k |k\rangle $ with energy $\varepsilon_k = \hbar^2 k^2/2m + |\mu|$ and wave function $|k\rangle = (u_k, -v_k )^T$ given by ~\cite{kaup90,kovrizhin01} 
\begin{equation}
\begin{pmatrix}u_k \\ - v_k\end{pmatrix} =
\frac{e^{ikx}}{(k^2\xi^2+1)}
\begin{pmatrix}
e^{-i \theta} [k\xi + i \tanh(x/\xi)]^2 \\
- e^{i \theta} {\rm sech}^2 (x/\xi)
\end{pmatrix} . \label{eq:bdgexcitation}
\end{equation}
Here we assume $N$ and $\mu$ to be given. Formally, BdG equations have eigenvalues with negative energies $-\varepsilon_k$ and wave functions $\overline{|k\rangle}=(-v_k^* , u_k^*)^T$. In addition, there are two zeromodes  given by $|\theta\rangle  = (\phi_0,-\phi_0^*)$ and $|X\rangle = - \xi \partial_x (\phi_0, \phi_0^*)$, corresponding to a small change in the phase and the soliton position, respectively. In the following, we neglect the phase degree of freedom since we are interested only in the soliton dynamics. The zeromodes cannot be treated as a small perturbation. The correct way to treat them nonperturbatively is via the collective coordinate method~\cite{rajamaran89,dziarmaga04}. 

Since the zeromodes have vanishing norm, they cannot  be included in the basis set that diagonalizes $K_{\rm BdG}$. Instead, the space of BdG excitations is supplemented by adjoint modes with non-zero norm, chosen as
\begin{equation}
K_\mathrm{BdG}|X^\mathrm{a}\rangle = \frac{\hbar^2} {M \xi^2} \sigma_3 |X\rangle ,
\end{equation}
where the mass $M$ is chosen such that the adjoint modes have unit overlap with the corresponding zeromodes. For the zeromode of soliton spatial translations, the adjoint mode is $|X^\mathrm{a}\rangle= - x/N \xi ~(\phi_0,- \phi_0^*)$ with mass $M = m N$. 
Now we promote the soliton coordinate to a quantum dynamical variable $X(t)$ and present the bosonic field $\phi(x,t)$ in terms of the complete basis set (quasiparticle eigenmodes and the adjoint to the zeromode of translations) as follows:
\begin{equation}
\begin{split}
\phi(x,t)=\phi_0(x- X(t))+ i \frac{\xi \pi_\mathrm{0}}{\hbar} u_X^a (x- X(t)) +\\ +
\sum_{k} \bigl[c_k(t) u_k (x- X(t)) - c_k^*(t) v_k^*(x- X(t))\bigr].
\label{dphi2}
\end{split}
\end{equation}
Here $\pi_\mathrm{0}$ is the bare momentum of the soliton without Bogoliubov quasiparticles. After substitution of (\ref{dphi2}) to the original Lagrangian~(\ref{eq:lagrangian}) and integrating out $\pi_\mathrm{0}$ we get~\cite{SM}
\begin{equation}
L=  
\frac{M \dot{X}^2}{2} + \pi_\mathrm{qp} \dot{X}+ \sum_k c_k^* [i \hbar \partial_t -\epsilon_k ] c_k.
\label{LagrangianSoliton}
\end{equation}
Here $\pi_\mathrm{qp}$ is the total momentum of Bogoliubov quasiparticles, while the momentum of soliton in their presence is given by $\pi_\mathrm{s}=M \dot{X}+\pi_\mathrm{qp}$. The explicit form of $\pi_\mathrm{qp}$ is given by 
\begin{equation}
\pi_\mathrm{qp}=\frac{1}{2}\sum_{k,k'}(c_k^*,c_k)\begin{pmatrix} \langle k| \sigma_z \hat{p} | k'\rangle & -\langle k| \sigma_z \hat{p} \overline{| k'\rangle} \\ -\overline{\langle k|} \sigma_z \hat{p} | k'\rangle & \overline{\langle k|} \sigma_z \hat{p} \overline{| k'\rangle} \end{pmatrix} \begin{pmatrix} c_{k'} \\ c_{k'}^* \end{pmatrix}
\label{MatrixElements}
\end{equation}   
where $\hat{p}=- i\hbar \partial_x$ is the momentum operator. Diagonal components $\pi^{\rm sc}$ correspond to scattering of quasiparticles, while nondiagonal components $\pi^{\rm ac}$ correspond to their emission and absorption. Using the explicit form of wave functions~(\ref{eq:bdgexcitation}) they can be found as follows 
\begin{eqnarray}
 \pi^{\rm sc}_{k^\prime k}=\frac{\pi\hbar}{3 \xi}\frac{({k^2-k^\prime}^2) ( {k^\prime}^2 + k^\prime k + k^2 + k_\xi^2)}{({k^\prime}^2+k_\xi^2) (k^2+k_\xi^2) \sinh\left[\frac{\pi}{2} \xi (k^\prime-k)\right]},\phantom{AAA} \label{Wd} \\ 
\pi^{\rm ac}_{k^\prime k}=\frac{\pi\hbar}{3\xi}\frac{(k+k^\prime)^2 ( {k^\prime}^2 - k^\prime k + k^2 + k_\xi^2)}{({k^\prime}^2+k_\xi^2) (k^2+k_\xi^2) \sinh\left[\frac{\pi}{2} \xi (k^\prime+k)\right]}, \phantom{AAA} \label{Wo}
\end{eqnarray} 
where we introduce the wavevector scale $k_\mathrm{\xi}=\xi^{-1}$. It should be noted that due to the integrability of the original problem the backscattering is suppressed $\pi^{\rm sc}_{k,-k}=0$. 

The Lagrangian~(\ref{LagrangianSoliton}) describes a motion of the soliton and Bogoliubov quasiparticles coupled with each other. The coupling term, $L_\mathrm{int}=\pi_\mathrm{qp} \dot{X}$, is new and important result of our work. The soliton is a quasiclassical entity while Bogoliubov quasiparticles can be treated as a quantum bath. The coupling with the bath leads to the friction and Langevin force in EOM of the soliton~(\ref{eq:EqOfMotion}). To derive such quasiclassical dissipative dynamics is an old, fundamental problem, which arises in the context of Brownian motion and the general Caldeira-Leggett model~\cite{caldeira83b}. However, since the coupling of the collective soliton coordinate to the bath is quadratic here, the problem at hand is more complicated than the Caldeira-Leggett model (where the coupling to the bath is linear and the model is exactly solvable, see also Refs.~\cite{CastroNeto1991, CastroNeto1993}).
We have derived the  quasiclassical  EOM  as the saddle point of a one-loop effective action in the Keldysh formalism. Formally, this corresponds to an expansion of the full action in terms of the soliton velocity,  $\dot{X}/c \ll 1$, where $c=\hbar/m\xi$ is a characteristic velocity scale in the model. Detailed technical calculations are presented in Supplemental Material~\cite{SM}, while here we present the main results -- Eq.~(\ref{eq:EqOfMotion}), which represents the quasiclassical Langevin EOM for the soliton. It is in effect Newton's second law for the soliton in the potential $U(X)$, supplemented with a retarded  friction force, $F[X(t)] = - \int_{0}^t dt' \eta(t-t') \dot{X}(t')$  and a stochastic  force, $f_s(t)$. The dissipation kernel $\eta(t)$  and correlation function $\langle f_s(t) f_s(0) \rangle$  are related via the  fluctuation-dissipation theorem and are expressed (see Eqs.~\eqref{eq:FrictionKernel} and~\eqref{eq:FrictionKernel2}) through the same spectral function $J(\omega)$ given by
\begin{align}
&J(\omega)=  2\pi \sum_{k k'} \Bigl[\underbrace{ 2 |\pi_{k k'}^{\rm sc}|^2 (f_{k'} - f_k) (\varepsilon_{k k'}^{-})^2 \delta(\varepsilon_{k k'}^- - \hbar\omega)}_{\mbox{scattering processes}} \nonumber \\
&+ \underbrace{|\pi_{k k'}^{\rm ac}|^2 (1 + f_{k'} + f_{k}) (\varepsilon_{k k'}^{+})^2 \delta(\varepsilon_{k k'}^{+} - \hbar \omega)}_{\mbox{annihilation/creation processes}} \Bigr], \label{eq:SpectralJ}
\end{align}
where $\varepsilon_{k k'}^{\pm} = \varepsilon_k \pm \varepsilon_{k'}$ and $f_k=f_\mathrm{B}(\varepsilon_k)$ is the Bose-Einstein distribution. 
The first term $J_\mathrm{sc}$ in (\ref{eq:SpectralJ}) corresponds to the scattering of Bogoliubov quasiparticles, while the second term $J_\mathrm{ac}$ originates from their annihilation/creation processes. Their dependencies on frequency are presented in Fig.~\ref{Fig:sf}. The $\omega$-dependence of $J_\mathrm{ac}$ weakly depends on temperature $T$ and has a threshold $2|\mu|/\hbar$, which is the minimal energy to create a pair of Bogoliubov quasiparticle in the superfluid of attractive bosons. At low frequencies \emph{only} $J_\mathrm{sc}$ survives. It does not have an Ohmic component (which would be linear in frequency), but is super-Ohmic
\begin{align}
\label{eq:SpectralJMarkovian}
J_\mathrm{M}(\omega) = \frac{16}{9\pi} \frac{m\hbar \omega^3}{|\mu|}e^{-\frac{|\mu|}{T}} = J_0 (\omega \tau)^3 ,
\end{align}
where $\tau=\hbar/|\mu|$ and $J_0=16 m/9\pi\tau^2 \exp[-|\mu|/T]$. Note that a super-Ohmic spectral function also appears for a impurity embedded to a bosonic superfluid~\cite{Peotta2013, Wadkin2012}. To get better insight into the origin of this effect (absence of Ohmic friction), rewrite the linear-in-$\omega$ part of the spectral function $J_\mathrm{sc}(\omega)$ as follows:
\begin{align}
J_\mathrm{sc}(\omega)&=\int d\varepsilon\, \nu(\varepsilon) \nu(\varepsilon+\omega) \left[f_\mathrm{B}(\varepsilon)-f_\mathrm{B}(\varepsilon+\omega)\right]\times  \nonumber \\ \times  S(\varepsilon+\omega,\varepsilon)
&\approx \omega \int_{-\infty}^\infty d\varepsilon \nu^2(\varepsilon) S(\varepsilon,\varepsilon) \left(-\frac{\partial f_\mathrm{B}}{\partial \varepsilon}\right), \label{eq:SpectralJLinear}  
\end{align} 
where $\nu(\varepsilon)=1/\pi\xi\sqrt{|\mu|(\epsilon-|\mu|)}$ is the density of states of Bogoliubov quasiparticles and we introduced
\begin{equation*}
S=\frac{2 \pi}{\nu(\varepsilon_1)  \nu(\varepsilon_2) }  \sum_{k' k} |\pi^\mathrm{sc}_{k k'}|^2 (\epsilon_{k,k'}^-)^2 \delta(\epsilon_{k} -\varepsilon_1)\delta(\epsilon_{k'} - \varepsilon_2),
\end{equation*}
which can be interpreted as the probability of  scattering of quasiparticles with energies $\varepsilon_1$ and $\varepsilon_2$.
From Eq.~(\ref{eq:SpectralJLinear}), we see that Ohmic friction comes exclusively from elastic scattering, which in 1D is equivalent to backscattering. However, as can be seen from Eqs.~\eqref{Wd} integrability ensures that $\pi^\mathrm{sc}_{k,-k}=0$. Hence, backscattering is forbidden -- and there is no Ohmic friction.

The dynamics of the soliton at macroscopic time scales $t \gg \tau$ is determined by the low-frequency part of the Fourier transform of the dissipation kernel $\eta(\omega)$. Its low-frequency asymptotics $\omega \tau \rightarrow 0$ corresponds to Markovian approximation and \emph{local} in time EOM. The real part $\eta'(\omega)=J(|\omega|)/|\omega|\approx J_0 \omega^2 \tau^3$, which is even and breaks time-reversal invariance at the quasiclassical level, is responsible for friction and originates from the scattering contribution to the spectral function $J_\mathrm{sc}$. The imaginary part of the dissipation kernel  $\eta''(\omega)=-i \delta M \omega$ is odd in frequency and is responsible for the mass renormalization $M\rightarrow M+\delta M$. The mass renormalization, however, is small with $\delta M/M \sim N^{-1}$ and can be neglected. The resulting EOM of the soliton in a trap with frequency $\omega_\mathrm{t}$ is in the Markovian approximation given by 
\begin{equation}
\label{eq:ALForce}
\ddot{X}- \tau_\mathrm{AL} \dddot{X}+\omega_{\rm t}^2 X  =  f_\mathrm{s}(t)/{M},
\end{equation}
The term $F[X] = \tau_\mathrm{AL} M \dddot{X}$ (with $\tau_\mathrm{AL}=J_0 \tau^3/M\approx 16 \tau/9\pi N$ here) can be recognized as the Abraham-Lorentz friction force, originally derived in the context of electrodynamics (where it describes the back reaction of electromagnetic radiation emitted by a charged particle on its motion). In our context,  pairs of Bogoliubov quasiparticles play the role of the radiation. The Abraham-Lorentz equation is plagued by spurious ``run away'' solutions and its regularization has been a controversial and long-standing problem \cite{LandauLifshitzII, Feynman1945}. The equation violates causality, which can be seen by calculating the response function 
$\chi^{-1}(\omega)=M_{\rm}[\omega_\mathrm{t}^2- \omega^2 (1+i\tau _\mathrm{AL}\omega)]$ of the soliton coordinate to the stochastic/external force $X(\omega)=\chi(\omega) f_s(\omega)$. The response  function is supposed to be analytical in the upper half-plane, while it has the spurious pole $\omega_\mathrm{AL}\approx i\tau_\mathrm{AL}^{-1}$, as is depicted in Fig.~2(b).

\begin{figure}[t]
\label{Fig2}
\includegraphics{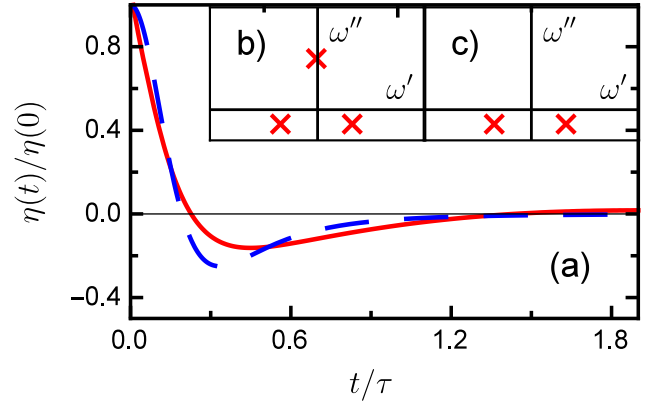}
\caption{(Color online) a) Time dependence of the friction kernel $\eta(t)$, corresponding to the exact spectral function, given by (\ref{eq:SpectralJ}), (red) and to the approximate one, given by  $J_\mathrm{M}^*=J_0 (\omega \tau)^3 \exp[-\omega \tau_* ]$ (dashed blue), at $T=\mu$ and $\tau_*=3 \tau $. b) Poles in complex the plane $\omega=\omega'+i\omega''$ of the response function $\chi(\omega)$ in the Markovian approximation leading to the Abraham-Lorentz friction force. In the upper half-plane there is the spurious pole corresponding to causality violation. c) Poles of the response function $\chi(\omega)$ calculated exactly for the spectral function $J_\mathrm{M}^*$ or in the Markovian approximation, if the Abraham-Lorentz force is treated as perturbation. Note that there is no unphysical pole in the upper half-plane of $\omega$.}
\vspace{-0.3 cm} 
\end{figure}

Note as a crucial point that the breaking of causality is an artifact of the Markovian approximation. Indeed, the location of the unphysical pole $|\omega_\mathrm{AL}|\tau\approx 9 \pi N_\mathrm{s}/16\gg 1$ is beyond the applicability of the Markovian approximation, which requires $\omega \tau \ll 1$. For regularization consider the spectral function  $J_\mathrm{M}^*(\omega)=J_0 (\omega \tau)^3 e^{- \omega \tau_*}$, where we parametrize its high-frequency part by a memory time $\tau_*$. The spectral function qualitatively captures the time dependence of the exact dissipation kernel, as is presented in Fig.~2(a). The exact expression for the response function $\chi(\omega)$, corresponding to the spectral function $J_\mathrm{M}^*(\omega)$ is given by $\chi^{-1}(\omega)=M \{ \omega_\mathrm{t}^2-\omega^2 - i \tau_\mathrm{AL}\omega^3 [\cosh(\omega \tau_*) + 2 \pi^{-1}  \cosh(\omega \tau_*)\mathrm{Si}(i \omega \tau_*)- 2 i \pi^{-1} \sinh (\omega \tau_*)  \mathrm{Ci}(-i \omega \tau_*) ] \},$ where $\mathrm{Ci}(-i\omega \tau_*)$ and $\mathrm{Si}( i\omega \tau_*)$ are cosine and sine integral functions. The response function is analytical in the upper half-plane, which ensures causality. 
Note also that the Abraham-Lorentz equation can be regularized by treating super-Ohmic friction as a perturbation, which leads to  $\ddot{X} + \tau_\mathrm{AL}\omega_\mathrm{t}^2 \dot{X} +\omega_\mathrm{t}^2 X = (f_\mathrm{s}+\tau_\mathrm{AL} \dot{f_\mathrm{s}})/{M}$.
The resulting response function is analytical in the upper half-plane with poles  $\omega \approx \pm \omega_\mathrm{t}-i \tau_{\mathrm{AL}} \omega_\mathrm{t}^2 $ (see Fig.~2(c)). In the presence of a trap potential the Abraham-Lorentz friction is well approximated by the usual friction $F[X]=- \tau_\mathrm{AL}\omega_\mathrm{t}^2 M \dot{X}$. Most importantly, this implies that the effective friction force is very sensitive to the trap frequency $\omega_\mathrm{t}$ and can be distinguished in this way from regular (extrinsic) Ohmic friction, which appears due to the breaking of integrability~\cite{Sinha2006}.   

In realistic experiments, the Abraham-Lorentz friction competes with the usual Ohmic friction. The Ohmic friction appears due to the quasi-one-dimensional nature of the trapping potential~\cite{Sinha2006}, or due to interactions between Bogoliubov quasiparticles~\cite{Gangardt2010}. However, the latter is a two-particle effect and is proportional to $\mathrm{exp}[-2|\mu|/T]$ instead of  $\mathrm{exp}[-|\mu|/T]$, which makes it unimportant at low temperatures. For estimates, we use the following parameters corresponding to the experiments in Ref.~\cite{Khaykovich2002}: particle number in the soliton $N \approx 1.5\; 10^3$; coherence length $\xi\approx 1.7 ~ \mathrm{\mu m}$;  chemical potential and temperature $|\mu|\approx 2 T \approx 11 ~\hbox {nK}$; axial trap frequency $\omega_\mathrm{t}\approx 2\pi \times 70 ~\hbox{Hz}$; transverse confinement and scattering length, $l_\perp=1.4 ~\mathrm{\mu m}$ and $a=0.44~\hbox{nm}$. The resulting decay times of the soliton are of comparable magnitude and given by $\tau_\mathrm{s}^\mathrm{AL}=9 \pi N  \tau  \mathrm{exp}[|\mu|/T]/16 \pi (\omega_\mathrm{t} \tau)^{2}\approx 187 ~\hbox{s}$ and $\tau_\mathrm{s}^{\mathrm{O}}\approx \pi N \tau \mathrm{exp}[|\mu|/T] (|\mu| x^2+T)^2/2 T \mu^2 x^4\approx 400 ~\hbox{s}$, where $\tau = \hbar/|\mu|$ and $x=N a/l_\perp$. Nevertheless, two mechanisms of friction can be distinguished in an experiment because $\tau_s^\mathrm{AL}\sim \omega_\mathrm{t}^{-2}$ strongly depends on axial trap frequency, while $\tau_s^\mathrm{O}$ is sensitive only to the transverse confinement $\omega_\perp$. This implies that the quantum soliton friction predicted in this work is within current experimental capabilities.

\begin{acknowledgments}
D.E. and V.G. are supported by US-ARO (contract \#W911NF1310172) and Simons Foundation. J.H. is supported by LPS-MPO-CMTC, JQI-NSF-PFC, ARO-Atomtronics-MURI, and Gonville and Caius College. The authors are grateful to Lev Pitaevskii
for illuminating discussions and a number of useful
suggestions.
\end{acknowledgments}

\bibliography{bib}

\widetext
\clearpage
\begin{center}
\textbf{\large Supplemental Material: "Non-Markovian quantum friction of bright solitons in superfluids" by Dmitry K. Efimkin, Johannes Hofmann, and Victor Galitski} 
\end{center}

\setcounter{equation}{0}
\setcounter{figure}{0}
\setcounter{table}{0}
\setcounter{page}{1}
\makeatletter
\renewcommand{\theequation}{S\arabic{equation}}
\renewcommand{\thefigure}{S\arabic{figure}}
\renewcommand{\bibnumfmt}[1]{[S#1]}
\renewcommand{\citenumfont}[1]{S#1}
\section{A. Quantization of zero-energy modes}
In the following, expanding on the discussion in the main text, we derive the coupling between soliton and Bogoliubov quasiparticles using the collective coordinate approach~\cite{SMrajamaran89}. Our starting point is the Lagrangian for a bosonic field 
\begin{equation}
\label{LagrangianBasic}
L = \int dx \left[\phi^* i \hbar \partial_t \phi - \frac{\hbar^2}{2m} |\nabla \phi|^2+\mu |\phi|^2-\frac{g_1}{2} |\phi|^4 \right]. 
\end{equation}
We introduce the soliton coordinate $X(t)$ as a quantum dynamical variable to substitute the zero energy mode, which corresponds to the soliton translation and requires a nonperturbative treatment. Next we will present the solution in the form $\phi(x,t)=\phi_0(x-X(t))+\delta \phi(x-X(t),t)$, where $\phi_0(x)$ is the solitonic solution and $\delta \phi(x-X(t),t)$ are quadratic fluctuations on top of the soliton, excluding translational zeromodes.

At first, we review the spectrum of quadratic fluctuations satisfying the BdG equations  
\begin{equation*}
\begin{pmatrix}
- \frac{\hbar^2 \nabla^2}{2m} - \mu + 2 g_1 |\phi_0|^2 & g_1 \phi_0^2 \\
g_1 \phi_0^{*2} & - \frac{\hbar^2 \nabla^2}{2m} - \mu + 2 g_1 |\phi_0|^2
\end{pmatrix}
\begin{pmatrix}
\delta\phi \\ \delta\phi^*
\end{pmatrix}=
\varepsilon \sigma_z \begin{pmatrix}
\delta\phi \\ \delta\phi^*
\end{pmatrix}.
\end{equation*}
The spectrum consists of Bogoliubov quasiparticles $|k\rangle=(u_k, -v_k)^T$, given by Eq.~(7) in the main paper, with positive energies $\varepsilon_k=\hbar^2 k^2/2m+|\mu|$, which are normalized as $\langle k|\sigma_z |k'\rangle=\delta_{k,k'}$. In addition, the BdG equations have solutions for states with opposite energies $\varepsilon_k=-\hbar^2 k^2/2m-|\mu|$ and wave function $\overline{|k\rangle}=(-v_k^* , u_k^*)^T$, which are normalized as $\overline{\langle k}|\sigma_z \overline{|k'\rangle}=-\delta_{k,k'}$. The spectrum of quadratic fluctuations also contains zeromodes  given by $|\theta \rangle = (\phi_0,-\phi_0^*)$ and $|X\rangle = - \xi \partial_x (\phi_0, \phi_0^*)$, corresponding to a small change in the phase and the soliton position, respectively. Their appearance is dictated by the presence of the translational and phase rotation symmetry of the condensate. Below we ignore $|\theta\rangle$, which is not important here. It is crucial, however, to describe phase diffusion in bosonics superfluids~\cite{SMLewenstein1996, SMImamoglu1997}.

Since the zeromode $|X \rangle$ has vanishing norm $\langle X| \sigma_z |X \rangle=0$, it needs to be supplemented by the adjoint mode $|X^a \rangle=(u_{X}^a, - v_{X}^a) = - x/N_0 \xi ~(\phi_0,- \phi_0^*)$, which also has vanishing norm $\langle X^a| \sigma_z| X^a \rangle=0$,  to form the complete basis for $K_\mathrm{BdG}$. The adjointness implies $K_\mathrm{BdG} | X^a \rangle =\hbar^2/ M \xi^2 \cdot \sigma_z | X\rangle $ and we chose $M=m N$ to have unit overlap with the mode translational zero mode, $\langle X^a| \sigma_z| X\rangle=1$. Below, we show that the mass $M$ is actually the inertial mass of the soliton. It should be noted that $|X \rangle\sim N^{1/2}$, while $|X^a \rangle\sim N^{-1/2}$, where the number of particles  $N$ in the soliton is large.

At first it is instructive to ignore the presence of Bogoliubov quasiparticles and present the bosonic field as follows 
\begin{equation}
\label{Ansatz1}
\phi(x,t) = \phi_0(x-X(t)) + i \frac{\xi \pi_\mathrm{0}}{\hbar} u_X^a (x-X(t)) .
\end{equation}
Here, the soliton coordinate $X(t)$ is treated as a dynamical degree of freedom, while $\pi_\mathrm{0}$ is supposed to be time independent and has dimension of momentum. After substituting (\ref{Ansatz1}) in the Lagrangian~(\ref{LagrangianBasic}) and integration out $\pi_0$, we get  
\begin{equation}
\label{LagrangianBare}
L_\mathrm{s}^0=\pi_\mathrm{0}\dot{X}-\frac{\pi_\mathrm{0}^2}{2 M}-E_0\rightarrow \frac{M \dot{X}^2}{2} - E_0. 
\end{equation}
Here, $M=m N$ is the mass introduced above and the last term $E_0=-N\hbar^2/3 m \xi^2$ is the energy of the bright soliton at rest. We also notice that $\pi_\mathrm{0}$ is actually the momentum of the soliton in the absence of Bogoliubov quasiparticles since $\pi_\mathrm{0}=\partial L /\partial \dot{X}$. Now, we reintroduce Bogoliubov quasiparticles and expand the bosonic field $\phi(x,t)$ as    
\begin{equation}
\label{Ansatz2}
\phi(x,t) = \phi_0(x-X(t)) + \sum_{k} \bigl[c_k(t) u_k (x-X(t)) - c_k^*(t) v_k^*(x-X(t))\bigr]+ i \frac{\xi \pi_\mathrm{0}}{\hbar} u_x^a (x-X(t)) .
\end{equation}
After substituting (\ref{Ansatz2}) into the Lagrangian~ (\ref{LagrangianBasic}) and integration out $\pi_0$, we get the coupling between the soliton velocity $\dot{X}$ and the $\emph{total}$ momentum of Bogoliubov quasiparticles $\pi_\mathrm{qp}$:
\begin{equation}
\label{LagrangianQP}
L_\mathrm{s}=\pi_\mathrm{0}\dot{X}-\frac{\pi_\mathrm{0}^2}{2 M }-E_0 + \pi_\mathrm{qp} \dot{X} +  \sum_k c_k^* [i \hbar \partial_t -\epsilon_k ] c_k  \rightarrow \frac{M \dot{X}^2}{2} - E_0 + \pi_\mathrm{qp} \dot{X}+ \sum_k c_k^* [i \hbar \partial_t -\epsilon_k ] c_k. 
\end{equation}
Here, we also neglected two other small terms which are proportional to $N^{-1/2}$, while all resulting terms are at least of the order of unity. We also note that the momentum of the soliton $\pi_\mathrm{s}=\partial L /\partial \dot{X}=\pi_\mathrm{0}+\pi_\mathrm{qp}$ is not just $\pi_\mathrm{0}$ but includes the contribution of quasiparticles. The explicit form of $\pi_\mathrm{qp}$ is
\begin{equation}
\pi_\mathrm{qp}=\frac{1}{2}\sum_{k,k'}(c_k^*,c_k)\begin{pmatrix} \langle k| \sigma_z \hat{p} | k'\rangle & -\langle k| \sigma_z \hat{p} \overline{| k'\rangle} \\ -\overline{\langle k|} \sigma_z \hat{p} | k'\rangle & \overline{\langle k|} \sigma_z \hat{p} \overline{| k'\rangle} \end{pmatrix} \begin{pmatrix} c_{k'} \\ c_{k'}^* \end{pmatrix}\equiv\frac{1}{2}\sum_{k,k'}(c_k^*,c_k)\begin{pmatrix} \pi_{k,k'}^\mathrm{sc} & \pi_{k,k'}^\mathrm{ac} \\ \pi_{k,k'}^\mathrm{ac} & \pi_{k,k'}^\mathrm{sc} \end{pmatrix} \begin{pmatrix} c_{k'} \\ c_{k'}^* \end{pmatrix}, 
\end{equation}
where $\hat{p}=- i\hbar \partial_x$ is the momentum operator and 
\begin{eqnarray}
 \pi^{\rm sc}_{k^\prime k}=\frac{\pi\hbar}{3 \xi}\frac{({k^2-k^\prime}^2) ( {k^\prime}^2 + k^\prime k + k^2 + k_\xi^2)}{({k^\prime}^2+k_\xi^2) (k^2+k_\xi^2) \sinh\left[\frac{\pi}{2} \xi (k^\prime-k)\right]},\phantom{AAA} \label{Wd} \quad 
\pi^{\rm ac}_{k^\prime k}=\frac{\pi\hbar}{3\xi}\frac{(k+k^\prime)^2 ( {k^\prime}^2 - k^\prime k + k^2 + k_\xi^2)}{({k^\prime}^2+k_\xi^2) (k^2+k_\xi^2) \sinh\left[\frac{\pi}{2} \xi (k^\prime+k)\right]}, \phantom{AAA} \label{Wo}
\end{eqnarray} 
where we introduce the wavevector scale $k_\mathrm{\xi}=\xi^{-1}$. Note that translational symmetry is broken by the presence of the soliton and the momentum operator is not diagonal in the eigenvalue basis of $K_\mathrm{BdG}$. The Hamiltonian of the system, corresponding to the Lagrangian (\ref{LagrangianQP}) can be rewritten as  
\begin{equation}
\label{Hamiltonian}
H_\mathrm{s}=\frac{(\pi_\mathrm{s}-\pi_\mathrm{qp})^2}{2 M}+\sum_k \varepsilon_k  c_k^* c_k.
\end{equation}
A similar coupling with quasiparticles appears for solitons in quantum field theory~\cite{SMCastroNeto1991, SMCastroNeto1993} and also is known in polaron dynamics~\cite{SMCastroNeto1992, SMTurkevich1987}.  
 
\section{B. Derivation of the equation of motion for a soliton}
Here, we derive the equation of motion for a soliton, (\ref{eq:langevin}), coupled with Bogoliubov quasiparticles. The key idea is the following: the complete system can be decomposed in two parts, the soliton (described by the collective position coordinate $X$) and its Bogoliubov excitations, which play the role of a bath. Integrating out the Bogoliubov excitations within a Keldysh formalism~\cite{SMKeldysh1964, SMAtlandSimons, SMKamenev} leads to an effective theory for the soliton coordinate with dissipation and Langevin noise. This is the same idea that underpins the analysis of the famous Caldeira-Leggett model~\cite{SMcaldeira83a, SMcaldeira83b}. It should be also noted that the derivation is not restricted to a Keldysh framework, another equivalent method is the Feynman-Vernon formalism~\cite{SMfeynman63, SMCastroNeto1991, SMCastroNeto1993}.

Our starting point is the Lagrangian describing the soliton coupled with Bogoliubov quasiparticles, which is derived in the main text of the paper, Eq.~(13), and is given by
\begin{align}
\label{Lagrangian}
L= \frac{M\dot{X}^2}{2}-U(X)  + \sum_k  c_k^*(i\hbar\partial_t-\varepsilon_k) c_k  - \frac{\dot{X}}{2} \sum_{kk'} \left[ 2 \pi_{kk'}^\mathrm{sc} c_{k}^*  c_{k'}+ \pi_{kk'}^\mathrm{ac} c_{k} c_{k'} + \pi_{kk'}^\mathrm{ac} c_{k}^* c_{k'}^*\right] ,
\end{align}
where we introduce an external trapping potential $U(X)$. According to the general philosophy of the Keldysh approach, we  duplicate all degrees of freedom: $c \rightarrow c_+, c_-$ and $X \rightarrow X_+, X_-$. They are now defined on the Keldysh contour where the index $\pm$ corresponds to its upper and lower branches. Introducing $X_\mathrm{c}=(X_\mathrm{+}+X_\mathrm{-})/2$, $X_\mathrm{q}=X_\mathrm{+}-X_\mathrm{-}$, and  $c_\mathrm{c(q)}=(c_{+}\pm c_{-})/\sqrt{2}$ (which are dubbed the classical and quantum degrees of freedom), the resulting Keldysh action for the Lagrangian \eqref{Lagrangian} can be written as $S^\mathrm{K}=S_1^\mathrm{K}+S_2^\mathrm{K}$, where $S_1^\mathrm{K}$ and $S_2^\mathrm{K}$ are given by
\begin{align}
S_1^\mathrm{K} &= \int dt\left[M \dot{X}_\mathrm{q} (t) \dot{ X}_\mathrm{c} (t) - U\left(X_\mathrm{c}(t)+\frac{X_\mathrm{q}(t)}{2}\right)+ U\left(X_\mathrm{c}(t)-\frac{X_\mathrm{q}(t)}{2}\right) \right]  \\
S_2^\mathrm{K} &= \int dt \Bigl[\int dt' \sum_k \hat{c}^*_k (t) \hat{G}^{-1}_k \hat{c}_k(t') - \frac{1}{2} \sum_{k,k'}\left\{2 \pi_{kk'}^\mathrm{sc} \hat{c}^\dagger_k (t)\hat{X} \hat{c}_{k'} (t) + \pi_{kk'}^\mathrm{ac} \hat{c}_k^T (t)\hat{X} \hat{c}_{k'} (t) + \pi_{kk'}^\mathrm{ac} \hat{c}^\dagger_k (t)\hat{X} \hat{c}_{k'}^* (t)\right\}\Bigr].  
\end{align}
Here $\hat{c}=(c_\mathrm{c},c_\mathrm{q})$, $\hat{X}=X_\mathrm{q}+2\sigma_x X_\mathrm{c}$ and the matrix Green function of Bogoliubov quasiparticles is given by
\begin{align}
\label{GreenFunctions}
\hat{G}_k(t,t')=
\begin{pmatrix}
G^K_k(t,t') & G^\mathrm{R}_k(t,t') \\[0.5ex]
G^\mathrm{A}_k	(t,t') & 0
\end{pmatrix}
&=
\begin{pmatrix}
-i (1+2 f_k) e^{- i \varepsilon_k (t-t')} & -i \Theta(t-t') e^{- i \varepsilon_k (t-t')} \\[0.5ex]
+i \Theta(t'-t) e^{- i \varepsilon_k (t-t')} & 0
\end{pmatrix},
\end{align}
where $G^\mathrm{R}_k(t,t'), G^\mathrm{A}_k(t,t'), G^K_k(t,t')$ are  retarded, advanced and Keldysh Green functions. The latter contains information about the distribution function of Bogoliubov quasiparticles, which we assume to be the thermal Bose-Einstein one $f_k = f_\mathrm{B}(\epsilon_k)$. Integrating out Bogoliubov  quasiparticles in $S_2^\mathrm{K}$ in the one-loop approximation, which assumes the soliton motion to be slow, $\dot{X}/c \ll 1$, where $c=\hbar/m\xi$ is a characteristic velocity scale in the model, we get the effective action for the soliton
\begin{align}
S_{\rm eff}^K &= S_1^\mathrm{K} + \frac{1}{2}\int dt \int dt' \, \biggl\{ \dot{X}_\mathrm{c}(t) \Pi_{cq}(t,t') \dot{X}_\mathrm{q}(t') + \dot{X}_\mathrm{q}(t) \Pi_{qc}(t,t') \dot{X}_\mathrm{c}(t') + \dot{X}_\mathrm{q}(t) \Pi_{qq}(t,t')  \dot{X}_\mathrm{q}(t') \biggr\},
\end{align}
where the functions $\Pi(t,t')$ are given by 
\begin{equation}
\begin{split}
&\Pi_{qc}(t,t') = \Pi_{qc}(t',t)=\sum_{kk'} \Bigl(4 |\pi_{kk'}^{\rm sc}|^2 \Bigl[G_{k',t't}^\mathrm{A} G_{k,tt'}^\mathrm{K} + G_{k',tt'}^\mathrm{R} G_{k,t't}^\mathrm{K}+\Bigr] + 2 |\pi_{kk'}^{\rm ac}|^2 \Bigl[G_{k',tt'}^\mathrm{R} G_{k,tt'}^\mathrm{K} + G_{k',t't}^\mathrm{A} G_{k,t't}^\mathrm{K} \Bigr] \Bigr), \\
&\Pi_{qq}(t,t') = \Pi_{qq}(t',t)=\sum_{kk'} \Bigl(4 |\pi_{kk'}^{\rm sc}|^2 \Bigl[G_{k',t't}^\mathrm{K} G_{k,tt'}^\mathrm{K} + G_{k',t't}^\mathrm{R} G_{k,tt'}^\mathrm{A} + G_{k',t't}^\mathrm{A} G_{k,tt'}^\mathrm{R}\Bigr] + \\
&\quad \quad \quad \quad \quad \quad \quad \quad \quad \quad + |\pi_{kk'}^{\rm ac}|^2 \Bigl[2 G^K_{k',tt'} G^K_{k,tt'} + G^\mathrm{A}_{k',tt'} G^\mathrm{A}_{k,tt'} +G^\mathrm{A}_{k',t't} G^\mathrm{A}_{k,t't} + G^\mathrm{R}_{k',tt'} G^\mathrm{R}_{k,tt'} +G^\mathrm{R}_{k',t't} G^\mathrm{R}_{k,t't} \Bigr]  \Bigr).
\end{split}
\end{equation}
Note that the functions $\Pi(t,t')=\Pi(\Delta t)$ depend only on the time difference $\Delta t=t-t'$ as well as Green functions of quasiparticles (\ref{GreenFunctions}). Using the explicit form of latter we find
\begin{align}
&\Pi_\mathrm{qc}(\Delta t) =  4 \Theta(\Delta t) \sum_{kk'} \Bigl(2 |\pi_{kk'}^{\rm sc}|^2 \bigl[f_{k'} - f_{k}\bigr] \sin[\varepsilon_{k,k'}^- \Delta t]  + |\pi_{kk'}^{\rm ac}|^2 \bigl[1 + f_k + f_{k'}\bigr] \sin[\varepsilon_{k,k'}^+ \Delta t] \Bigr), \\&
\Pi_\mathrm{qq}(\Delta t) =  4 \sum_{kk'} \Bigl(2 |\pi_{kk'}^{\rm sc}|^2 \bigl[f_k + f_{k'} + 2 f_k f_{k'}\bigr] \cos[\varepsilon_{k',k}^- \Delta t] + 4 |\pi_{kk'}^{\rm ac}|^2 \bigl[1 + f_k + f_{k'} + 2 f_k f_{k'}\bigr] \cos[\varepsilon_{k',k}^+ \Delta t] \Bigr) .
\end{align}
Here we have introduced compact notations $\varepsilon_{k',k}^\pm=\varepsilon_{k'}\pm\varepsilon_{k}$. Expanding in quantum coordinate $X_\mathrm{q}$ and performing the integration by parts we get
\begin{align}
S_{\rm eff}^\mathrm{K} = \int dt \Biggl\{X_\mathrm{q}(t)\Bigl[ - M \ddot{X}_\mathrm{c}(t)-\partial_{X_\mathrm{c}} U\left(X_\mathrm{c}\right) \Bigl] + \int dt' \, \biggl[ -X_\mathrm{q}(t) \eta(t-t') \dot{X}_\mathrm{c}(t') + \frac{1}{2} X_\mathrm{q}(t) C_s(t-t') X_{q}(t') \biggl] \Biggr\} ,
\end{align}
where $\eta(t-t') = \partial_t \Pi_{\mathrm{qc}}(t,t')$ and $C_\mathrm{s}(t-t') = \partial_t \partial_{t'} \Pi_{\mathrm{qq}} (t,t')$. The last term can be decoupled with the help of a Hubbard-Stratonovich transformation with an auxiliary field $\xi$:
\begin{align}
\label{ActionHS}
S_{\rm eff}^\mathrm{K} = \int dt \Biggl\{X_\mathrm{q}(t)\Bigl[ -M \ddot{X}_\mathrm{c}(t)- \partial_{X_\mathrm{c}} U\left(X_\mathrm{c}\right)+\xi(t)- \int dt' \eta(t-t') \dot{X}_\mathrm{c}(t')\Bigl] +  \frac{1}{2} \int dt' \xi(t) C_s^{-1}(t-t') \xi(t') \Biggr\}.\end{align}
The saddle point of this action, which can be found by extremizing with respect to the quantum coordinate $X_\mathrm{q}$, gives the quasiclassical equation of motion for the soliton coordinate $X\equiv  X_\mathrm{c}$ as follows  
\begin{align}
M \ddot{X}(t) + \int_{0}^t dt' \, \eta(t-t') \dot{X}(t') &= -\partial_X U(X)+\xi(t) , \label{eq:langevin}
\end{align}
where $\eta (\Delta t)$ is the damping kernel. The auxiliary field $\xi(t)$ appears as Langevin noise; its autocorrealtion function $\langle \xi(t) \xi(t') \rangle$ can be deduced from \eqref{ActionHS} and is given by $\langle \xi(t) \xi(t') \rangle=C_\mathrm{s}(\Delta t)$. The explicit form of $\eta (t)$ and   $C_\mathrm{s}(t)$ is given by 
\begin{align}
&\eta(\Delta t)=  4 \Theta(t) \sum_{kk'} \Bigl(2 |\pi_{kk'}^{\rm sc}|^2 \varepsilon_{k,k'}^- \bigl[f_{k'} - f_{k}\bigr] \cos[\varepsilon_{k,k'}^- \Delta t] + |\pi_{kk'}^{\rm ac}|^2 \varepsilon_{k,k'}^+ \bigl[1 + f_{k'} + f_{k}\bigr] \cos[\varepsilon_{k,k'}^+ \Delta t] \Bigr), \\
& C_s(\Delta t) =   4 \sum_{kk'} \Bigl(2 |\pi_{kk'}^{\rm sc}|^2 \varepsilon_{k,k'}^{-,2} \bigl[f_k + f_{k'} + 2 f_k f_{k'}\bigr] \cos[\varepsilon_{k',k}^- \Delta t] + |\pi_{kk'}^{\rm ac}|^2 \varepsilon_{k,k'}^{+,2} \bigl[1 + f_{k'} + f_{k} + 2 f_{k'} f_{k}\bigr] \cos[\varepsilon_{k,k'}^+ \Delta t] \Bigr) .
\end{align}
Note that according to the fluctuation-dissipation theorem they are not independent and are intrinsically connected. Particularly, as can be proven by direct substitution, they can be written in the compact form 
\begin{equation}
\label{eq:FrictionKernel}
\eta(\Delta t) = \frac{2}{\pi}\int_0^\infty d\omega \, \frac{J(\omega)}{\omega} \cos \left(\omega \Delta t\right), \quad \quad \quad
C_\mathrm{s}(\Delta t)  = \frac{2\hbar }{\pi}\int_0^\infty d\omega \, J(\omega) \coth \left( \frac{\hbar \omega}{2 T} \right) \cos \left(\omega \Delta t\right).
\end{equation}
Here, $J(\omega)$ is the spectral function of the bath formed by Bogoliubov quasiparticles for a soliton, which is given by

\begin{equation}
J(\omega)=  2\pi \sum_{k k'} \Bigl[ 2 |\pi_{k k'}^{\rm sc}|^2 (f_{k'} - f_k) (\varepsilon_{k k'}^{-})^2 \delta(\varepsilon_{k k'}^- - \hbar\omega) + |\pi_{k k'}^{\rm ac}|^2 (1 + f_{k'} + f_{k}) (\varepsilon_{k k'}^{+})^2 \delta(\varepsilon_{k k'}^{+} - \hbar \omega) \Bigr].
\label{SpectralJ}
\end{equation}

\end{document}